\documentclass[prl,aps,twocolumn,10pt]{revtex4-1}
\usepackage[utf8x]{inputenc}
\usepackage{amsmath,amssymb,graphicx,xcolor,esint}

\newcommand{\CeigA}{\mathbf{C}^{\left( \alpha \right)}_{\substack{ e \\ o}   m n l} \left(\mathbf{r} \right)}
\newcommand{\CeigAe}{\mathbf{C}^{\left( \alpha \right)}_{e   m n l} \left(\mathbf{r} \right)}
\newcommand{\CeigAo}{\mathbf{C}^{\left( \alpha \right)}_{o m n l} \left(\mathbf{r} \right)}
\newcommand{\CeigB}{\mathbf{C}^{\left( \beta \right)}_{\substack{ e \\ o}  m n l} \left(\mathbf{r} \right)}
\newcommand{\CeigBe}{\mathbf{C}^{\left( \beta \right)}_{e   m n l} \left(\mathbf{r} \right)}
\newcommand{\CeigBo}{\mathbf{C}^{\left( \beta \right)}_{o   m  n l} \left(\mathbf{r} \right)}

\newcommand{\Eso}{{\bf \tilde{E}}_{S}^{\left(1\right)}}
\newcommand{\Esd}{{\bf \tilde{E}}_{S}^{\left(2\right)}}
\newcommand{\Esdtot}{{\bf E}_{S}^{\left(2\right)}}
\newcommand{\Est}{{\bf \tilde{E}}_{S}^{\left(3\right)}}
\newcommand{\Ei}{{\bf E}_{i}}
\newcommand{\Eit}{{\bf \tilde{E}}_{i}}

\newcommand{\Mi}[3]{{\bf M}_{#3}^{\left(#1\right)} \left( #2 \right)}
\newcommand{\Ni}[3]{{\bf N}_{#3}^{\left(#1\right)} \left( #2 \right)}
\newcommand{\Mii}[2]{{\bf M}_{#2}^{\left(#1\right)}}
\newcommand{\Nii}[2]{{\bf N}_{#2}^{\left(#1\right)}}

\newcommand{\re}[1]{\mbox{Re}\left\{#1\right\}}
\newcommand{\no}{\hat{\bf n}}
\newcommand{\im}[1]{\mbox{Im}\left\{#1\right\}}

\begin{document}

\title{Spectral theory of electromagnetic scattering by a coated sphere}

\author{Mariano Pascale}
\affiliation{ Department of Electrical Engineering and Information Technology, Universit\`{a} degli Studi di Napoli Federico II, via Claudio 21,
 Napoli, 80125, Italy}
\author{Giovanni Miano}
\affiliation{ Department of Electrical Engineering and Information Technology, Universit\`{a} degli Studi di Napoli Federico II, via Claudio 21,
 Napoli, 80125, Italy}
\author{Carlo Forestiere}
\affiliation{ Department of Electrical Engineering and Information Technology, Universit\`{a} degli Studi di Napoli Federico II, via Claudio 21,
 Napoli, 80125, Italy}

\begin{abstract}
In this paper, we introduce an alternative representation of the electromagnetic field scattered from a homogeneous sphere coated with a homogeneous layer of uniform thickness. Specifically, we expand the scattered field using a set of modes that are independent of the permittivity of the coating, while the expansion coefficients are simple rational functions of the permittivity. The theory we develop  represents both a framework for the analysis of plasmonic and photonic modes and a straightforward methodology to design the permittivity of the coating to pursue a  prescribed tailoring of the scattered field. To illustrate the practical implications of this method, we design the permittivity of the coating to zero either the backscattering or a prescribed multipolar order of the scattered field, and to maximize an electric field component in a given point of space.
\end{abstract}
\maketitle

In the last few years, the coated sphere has represented the ideal framework for the investigation of emerging physical phenomena at the nanoscale. In particular, it has been used to exemplify many properties of metal nanostructures, including the frequency tunability of the plasmon resonance \cite{oldenburg1998nanoengineering},  the plasmon hybridization  \cite{prodan2003hybridization}, and Fano-like resonant lineshapes \cite{luk2010fano}. In addition, coated spheres have inspired new devices such as scattering cancellation cloaks \cite{Alu05}, spaser-based nanolaser \cite{noginov2009demonstration}, and have been also used for the plasmon-enhanced molecular fluorescence \cite{tam2007plasmonic}, and for the imaging and therapy of cancer \cite{loo2005immunotargeted}.
 
Aden and Kerker \cite{aden1951scattering} first obtained the analytical solution of the problem of electromagnetic scattering from a homogeneous sphere coated with a homogeneous layer of uniform thickness. Subsequently, Li Kai and Massoli \cite{Kai:94} proposed an extension to multi-layers spherical particles. Over the years, several algorithms have been also developed to improve the efficiency and accuracy of the numerical solution  \cite{Toon:81,kaiser1993stable,wu1997improved}. However, the Mie theory and its extensions such as the one proposed by Aden and Kerker are not based on spectral theories. Specifically, vector spherical wave functions are not eigenmodes of any formulation of the Maxwell's equations in the presence of a coated sphere. 
A spectral theory can be of great use in the analysis of resonances and of anomalous scattering phenomena, such as Fano lineshapes \cite{luk2010fano}, because it allows one to rigorously identify the principal modes contributing to overall scattered field.

Moreover, in the Mie-Aden-Kerker solution, the contributions of the material parameters and of the geometry are mathematically intertwined and cannot be separated. Specifically, the expansion coefficients of the scattered field in terms of VSWFs are complicated functions of both the radius and the electric permittivity of the coating. Thus, the design of the cloak to achieve assigned constraints on the scattered electromagnetic field is usually cumbersome. For instance, although the design of the permittivity of the coating can be carried out analytically in the quasi-electrostatic limit \cite{Alu05}, and semi-analytically for particles of dimensions less than the incoming wavelength by using perturbation theory \cite{forestiere2014cloaking}, researchers have to resort to numerical optimization in the general case.

In this manuscript, we derive an alternative formulation of the scattering problem from a homogeneous sphere with permittivity $\varepsilon_{r\,1}$ coated with a homogeneous layer of uniform thickness and permittivity $\varepsilon_{r\,2}$, based on an auxiliary eigenvalue problem. The main feature of the proposed method is that the scattered electric field is represented through a series expansion, where the $s$-th addend has the form $\left( \gamma_s - \varepsilon_{r\,2} \right)^{-1} {\bf C}_s$,  where $\left\{ \gamma_s \right\}$ and $\left\{ {\bf C}_s \right\}$ are respectively the eigenvalues and the eigenvectors of an auxiliary eigenvalue problem defined in the following, which do not depend on the permittivity of the coating. This expansion enables the achievement of two goals. The identification of the dominant modes of the scattered electromagnetic field and the design of the permittivity of the coating $\varepsilon_{r\,2}$ to achieve a prescribed tailoring of the scattered field, exploiting the fact that the expansion coefficients of the scattered field are a rational function of $\varepsilon_{r\,2}$. This work represents the extension to the case of a coated sphere of the approach proposed in Ref. \cite{Forestiere16}, where it has been explicitly applied only to the case of a homogeneous sphere.  Our approach naturally leads to the one developed in Ref. \cite{mayergoyz2007numerical,roman2011designing} for a coated object in the quasi-electrostatic limit. Analogous formulations have been introduced in the past \cite{Bergman80} and applied to the quasi-static limit \cite{Bergman78,Bergman80,Fredkin2003,Mayergoyz05}, to the scalar Mie scattering \cite{Markel10}, and to describe the full-wave electromagnetic response of a flat-slab composite structure \cite{Bergman16}.

The paper is organized as follows. The differential formulation of the scattering problem from an arbitrary coated object is introduced in Sec. \ref{sec:Formulation}, together with the corresponding auxiliary eigenvalue problem. In this section, we also derive the main properties of its eigenvalues and eigenmodes, and we show how the scattered field can be represented in terms of eigenmodes which are independent of the material of the coating. Then, we devote Sec. \ref{sec:FormulationSphere}  to particularize these results to the case of a coated sphere, providing the expression of the characteristic polynomial and of the eigenmodes. Next, in Sec. \ref{sec:Analysis}	 we show how the introduced approach represents the natural framework for the analysis of plasmonic and photonic resonances in core-shell nanoparticle.  Eventually, in Sec \ref{sec:Design} we use the proposed approach to design the permittivity of the coating to tailor the scattered field in a prescribed way, exploiting the fact that the expansion coefficients are a rational function of the permittivity. We carry out several examples, designing the permittivity of the coating to zero the backscattering, to zero  a prescribed multipolar scattering order, and to maximize the electric field in a given point of space.

\section{General Formulation}
\label{sec:Formulation}
\begin{figure}
\centering
\includegraphics[width=\linewidth]{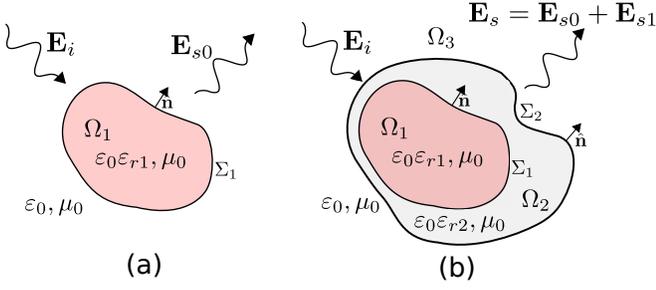}
\caption{Sketch of the two considered scenarios. (a) A homogeneous object of permittivity $\varepsilon_{r\,1}$. (b) The same object is then covered with a homogeneous layer of relative permittivity $\varepsilon_{r\,2}$. Both the systems are excited by the same incident field ${\bf E}_i$.
}
  \label{fig:Figure_1}
\end{figure}

Let us consider the electromagnetic scattering by an object occupying a regular region $\Omega_1$, shown in Fig. \ref{fig:Figure_1} (a). The object is excited by a time harmonic electromagnetic field incoming from infinity $\re{\Ei \left({\bf r}\right) e^{- i \omega t}}$. The material of the object is a non-magnetic isotropic homogeneous lossless dielectric with relative permittivity $\varepsilon_{r\,1}$, surrounded by vacuum. We denote the field scattered by the object as $ {\bf E}_{s \,0}$.
Now, in order to modify the scattering properties of this object, we cover the domain $\Omega_1$ with an arbitrarily shaped homogeneous coating as sketched in Fig. \ref{fig:Figure_1} (b). The coating is made of a linear, homogeneous, isotropic, time-dispersive material with relative permittivity $\varepsilon_{r \,2}$.  We denote with $\Omega_2$ the regular region occupied by the shell, and with $\Omega_3$ the
external space;  we also denote with $\Sigma_1$ and $\Sigma_2$ the surfaces separating the shell from the core and with the external space, respectively. The outward-pointing normals to the two surfaces $\Sigma_1$ and $\Sigma_2$ are both indicated with $\bf n$. The object is still excited by the field $\re{\Ei \left({\bf r}\right) e^{- i \omega t}}$. 

Let ${\bf E}_S^{\left(j\right)}$ be the scattered electric fields in $\Omega_j$, $\forall j \in \left\{ 1,2,3 \right\}$. It can be decomposed as 
\begin{equation}
 {\bf E}_S^{\left(j\right)} = \tilde{\bf E}_{S}^{\left(j\right)} +  {\bf E}_{S\,0}.
\end{equation}
The field $\tilde{\bf E}_{S}^{\left(j\right)}$ represents the change in the scattered field caused by the introduction of the coating. It is solution of the following problem:
\begin{alignat}{2}
\label{eq:MErot1}
& k_0^{-2} \boldsymbol{\nabla}^2 \Eso -  \varepsilon_{r \, 1}   \Eso = {\bf 0} && \mbox{in} \, \Omega_1, \\
\label{eq:MErot2}
& k_0^{-2} \boldsymbol{\nabla}^2 \Esd - \varepsilon_{r\,2} \left( \omega \right) \Esd =  \left[ \varepsilon_{r\,2} \left( \omega \right) -1  \right] \Eit  \quad && \mbox{in} \, \Omega_2, \\
\label{eq:MErot3}
& k_0^{-2} \boldsymbol{\nabla}^2 \Est - \Est = {\bf 0}  \; && \mbox{in} \, \Omega_3, 
\end{alignat}

\begin{equation}
\begin{aligned}
\label{eq:BC1}
   & \no \times \left( \Esd - \Eso \right) = {\bf 0} \\
   & \no \times \left( \boldsymbol{\nabla} \times \Esd - \boldsymbol{\nabla} \times\Eso \right) = {\bf 0}
   \end{aligned} \qquad
   \mbox{on} \, \Sigma_1, \\
\end{equation}
\begin{equation}
   \begin{aligned}
   \label{eq:BC2}
   & \no \times \left( \Est - \Esd \right) = {\bf 0} \\
   & \no \times \left( \boldsymbol{\nabla} \times \Est - \boldsymbol{\nabla} \times\Esd \right) = {\bf 0} \\
\end{aligned} \qquad   \mbox{on} \, \Sigma_2, \\
\end{equation}
where $k_0= \omega/c_0$, $c_0$ is the light velocity in vacuum, and
\begin{equation}
{\bf \tilde{E}}_i = \mathbf{E}_i + \mathbf{E}_{S\,0}.
\label{eq:EincDec}
\end{equation}
  Equations \ref{eq:MErot1}-\ref{eq:BC2} have to be solved with the  radiation conditions, namely the regularity and Silver-M\"uller conditions at infinity
\begin{equation}
\begin{aligned}
  & \Est + \frac{1}{i k_0} \hat{\bf r} \times  \boldsymbol{\nabla} \times \Est = o \left( \frac{1}{r} \right), \\
  & \Est = o \left( \frac{1}{r} \right), \quad  \boldsymbol{\nabla} \times \Est = o \left( \frac{1}{r} \right),
  \end{aligned}
  \label{eq:SilverMuller}
\end{equation}
which constraint the scattered field to be an outgoing wave.  This problem has a unique solution $\forall \im {\varepsilon_{r 2}} \ge 0 $ \cite{cessenat1996}.
Since our main goal is the study the behaviour of the solution as $\varepsilon_{r\,2}$ varies, we introduce the following auxiliary eigenvalue problem
\begin{alignat}{2}
\label{eq:AuxCalderonint}
& \hat{\bf n} \times \boldsymbol{\nabla} \times {\bf C}  = \mathcal{C}^{i} \left\{ \hat{\bf n} \times {\bf C} \right\} \qquad && \mbox{on} \; \Sigma_1, \\
\label{eq:AuxHelmholtz}
 & -{k_0^{-2}} \, \boldsymbol{\nabla}^2 {\bf C}  =  \gamma {\bf C} \qquad &&  \mbox{in} \; \Omega, \\
 \label{eq:AuxCalderonext}
& \hat{\bf n} \times \boldsymbol{\nabla} \times {\bf C}  = \mathcal{C}^{e} \left\{ \hat{\bf n} \times {\bf C} \right\} \qquad && \mbox{on} \; \Sigma_2,
\end{alignat}
where $\gamma$ is the eigenvalue and ${\bf C}\left( {\bf r}\right)$ is the corresponding eigenfunction. We introduced the exterior outgoing Calder\'on operator $\mathcal{C}^{e}$ \cite{cessenat1996} that takes the tangential component of the  field $\Est$ on $\Sigma_2$, i.e. $\left. \hat{\bf n} \times \Est \right|_{\Sigma_2} $,  and returns the tangential component of its curl $ \left. \hat{\bf n} \times \boldsymbol{\nabla} \times \Est \right|_{\Sigma_2} $, i.e.
\begin{equation}
  \mathcal{C}^{e} \left\{  \left. \hat{\bf n} \times \Est  \right|_{\Sigma_2} \right\} = \left. \no \times \boldsymbol{\nabla} \times \Est \right|_{\Sigma_2}.
\end{equation}
Analogously, we introduce the interior Calder\'on operator $\mathcal{C}^{i}$ \cite{cessenat1996} that takes the tangential component of the field $\Eso$ on $\Sigma_1$, i.e. $\left. \hat{\bf n} \times \Eso \right|_{\Sigma_1} $, and returns the tangential component of its curl $ \left. \hat{\bf n} \times \boldsymbol{\nabla} \times \Eso \right|_{\Sigma_1} $, namely:
\begin{equation}
  \mathcal{C}^{i} \left\{  \left. \hat{\bf n} \times \Eso  \right|_{\Sigma_1} \right\} = \left. \no \times \boldsymbol{\nabla} \times \Eso \right|_{\Sigma_1}.
\end{equation}
Equations \ref{eq:AuxCalderonint}, \ref{eq:AuxCalderonext} are  equivalent, respectively, to the set of equations \ref{eq:MErot1},\ref{eq:BC1} and to the set of equations \ref{eq:MErot3},\ref{eq:BC2},\ref{eq:SilverMuller}.
Since the operator $-\boldsymbol{\nabla}^2$ in $\Omega$ with the boundary conditions \ref{eq:AuxCalderonint},\ref{eq:AuxCalderonext} is compact, its spectrum $\left\{ \gamma_s \right\}_{s \in \mathbb{N}}$ is countably infinite. 
 This fact is a consequence of the radiation conditions, which are implicitly accounted for by the Calder\'on operator.
 
  In this case, the operator $-\boldsymbol{\nabla}^2$ is not Hermitian (even though symmetric), thus its eigenvalues $\gamma_s$ are complex with $\im {\gamma_s} < 0 $. The  eigenmodes ${\bf C}_s$ and ${\bf C}_r$ corresponding to different eigenvalues $\gamma_s$ and $\gamma_r$ are not orthogonal in the usual sense, i.e. $\langle {\bf C}_s^*,{\bf C}_r\rangle_\Omega \ne 0$, where
 \begin{equation}
\langle \mathbf{A},\mathbf{B} \rangle_V = \iiint_V \mathbf{A} \cdot \mathbf{B} \, \mbox{dV}.
\end{equation}
 Nevertheless, by introducing its dual eigenvalue problem   it can be proved that
\begin{equation}
 \langle {\bf C}_s,{\bf C}_r\rangle_{\Omega_2} = 0 \qquad  \gamma_{r} \ne \gamma_{s},
 \label{eq:Orthogonality}
\end{equation}
and
\begin{align}
 \re{\gamma_s} &= \frac{1}{\left\| {\bf C}_s \right\|^2_{\Omega_{2}}}  \left[\frac{\left\| \boldsymbol{\nabla} \times {\bf C}_s \right\|^2_{\mathbb{R}^3}}{k_0^2} -\varepsilon_1 \left\| {\bf C}_s \right\|^2_{\Omega_1}  -\left\| {\bf C}_s \right\|^2_{\Omega_3} \right], \label{eq:BalanceRe} \\
 \im{\gamma_s} &= - \frac{1}{\left\| {\bf C}_s \right\|^2_{\Omega_{2}}}   \varoiint_{S_\infty}   \frac{\left| {\bf C}_s \right|^2}{k_0} \mbox{dS},
 \label{eq:BalanceIm}
\end{align}
where $\left\| {\bf A} \right\|_{V}^2 = \langle {\bf A}^*, {\bf A} \rangle_{V}$. The eigenfunction ${\bf C}_s$ are extended in $\mathbb{R}^3$ by requiring that they satisfy Eq. \ref{eq:MErot1},\ref{eq:MErot3}, the boundary conditions \ref{eq:BC1}-\ref{eq:BC2} and the radiation conditions at infinity \ref{eq:SilverMuller}. 

 Equation \ref{eq:BalanceRe} suggests that $\re{\gamma_s}$ does not have a definite sign, while Eq. \ref{eq:BalanceRe} shows that $\im{\gamma_s}$ is strictly negative. In particular, $\im{\gamma_s}$ is proportional to the contribution of the corresponding eigenfunction to the power radiated to infinity, accounting for its radiative losses.

In the presence of an arbitrary external excitation $\Ei$, the solution of the scattering problem is
\begin{equation}
   \Esdtot \left( {\bf r} \right) = {\bf E}_{S\,0} \left( {\bf r} \right) + \left( \varepsilon_{r\,2} - 1 \right) \displaystyle \sum_{s=1}^\infty \frac{1}{\gamma_s - \varepsilon_{r \, 2}  } \frac{\langle {\bf C}_s, {\bf \tilde{E}}_i \rangle_{\Omega_2}}{\langle {\bf C}_s, {\bf C}_s \rangle_{\Omega_2}} {\bf C}_s \left( {\bf r} \right),
   \label{eq:Exp}
\end{equation}
where $\tilde{\mathbf{E}}_i \left( {\bf r} \right)$ is given by Eq. \ref{eq:EincDec}.
The eigenvalues $\gamma_s$ and the eigenfunctions ${\bf C}_s$ are  independent of the permittivity $\varepsilon_{r2}$, depending solely on the geometry of the coated object and on the permittivity of the core $\varepsilon_{r1}$. The permittivity $\varepsilon_{r2}$ appears in the multiplicative factors only as $ \left( \varepsilon_{r2} - 1 \right)/\left( \varepsilon_{r2} - \gamma_s \right)$.

\section{Coated Sphere}
\label{sec:FormulationSphere}
\begin{figure}
\centering
\includegraphics[width=40mm]{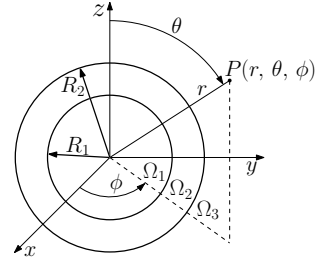}
\caption{Sketch of a homogeneous sphere coated with a homogeneous layer of uniform thickness.}
  \label{fig:CoatedSphere}
\end{figure}

From now on, we assume that the region $\Omega_1$ is a sphere of radius $R_1$, while the region $\Omega_2$ is a concentric layer with uniform thickness $R_2 - R_1$, as sketched in Fig. \ref{fig:CoatedSphere}. 
We define the dimensionless quantities
\begin{equation}
\begin{aligned}
x &= 2 \pi \, {R_1}/{\lambda}, \\
y &= 2 \pi \, {R_2}/{\lambda},
\end{aligned}
\end{equation}
 where $\lambda = 2 \pi c_0 / \omega$. We also introduce the aspect ratio $\eta$, as the ratio between the inner and the outer radius, $\eta = R_1 / R_2$. 
 
\subsection{Eigenvalues and Eigenfunctions} 
 The set of eigenvalues $\left\{ \gamma_s \right\}_{s \in \mathbb{N}}$ is the union of  $\left\{ \alpha_{nl} \right\}_{ \left( n,l \right) \in \mathbb{N}^2}$ and $\left\{ \beta_{nl} \right\}_{ \left( n,l \right) \in \mathbb{N}^2}$ being $\alpha_{nl}$ (respectively $\beta_{nl}$) the $l$-th root of the power series $\mathcal{P}_n$ (respectively $\mathcal{Q}_n$):
\begin{align}
   \mathcal{P}_n \left( \alpha \right) &= \sum_{h=0}^{\infty}\left(\alpha -1\right)^h\sum_{k=0}^{h}\sigma_{hk}\left[ r_{hk}^{\left( n \right)}\, \alpha^2 +  s_{hk}^{\left( n \right)}\, \alpha+ {t}_{hk}^{\left( n \right)} \right],
\label{eq:PolyP} \\
   \mathcal{Q}_n \left( \beta \right) &= 
\sum_{h=0}^{\infty}\left(\beta -1\right)^h\sum_{k=0}^{h}\sigma_{hk}\left[ u_{hk}^{\left( n \right)} \beta+ {v}_{hk}^{\left( n \right)} \right],
\label{eq:PolyQ}
\end{align}
where the expressions of coefficients $\sigma_{hk}$, $r_{hk}^{\left( n \right)},s_{hk}^{\left( n \right)},t_{hk}^{\left( n \right)}$ and $u_{hk}^{\left( n \right)},v_{hk}^{\left( n \right)}$ are given in the Appendix \ref{sec:PolyCoef}.
The eigenspace corresponding to the eigenvalue $\alpha_{nl}$ is spanned by the eigenfunctions $\CeigA$ with $m \in \mathbb{N}_0$ and $m \le n$, given by:
\begin{equation}
\CeigA=
 \begin{cases}
d_{nl} \mathbf{N}_{\substack{ e \\ o} m  n}^{(1)}( \sqrt{\varepsilon_{r\,1}} \, k_0 \mathbf{r}) &  \mathbf{r}\in \Omega_1, \\
\mathbf{N}_{\substack{ e \\ o}   m   n}^{(1)}( \sqrt{\alpha_{nl}} \, k_0 \mathbf{r})+ g_{nl} \mathbf{N}_{\substack{ e \\ o}   m   n}^{(2)}(\sqrt{\alpha_{nl}} \, k_0 \mathbf{r}) & \mathbf{r}\in \Omega_2, \\
-a_{nl} \mathbf{N}_{\substack{ e \\ o}   m   n}^{(3)}(k_0 \mathbf{r}) & \mathbf{r}\in \Omega_3,
\end{cases}
\label{eq:ElectricModes}
\end{equation}
They feature zero radial magnetic field. Therefore, they are denoted as {\it electric type} modes. 
The eigenspace associated to the eigenvalue $\beta_{nl}$ is spanned by the eigenfunctions $\CeigB$ with $m  \le n$, given by:
\begin{equation}
\CeigB=
 \begin{cases} c_{nl}
  \mathbf{M}_{\substack{ e \\ o}  m  n}^{(1)}( \sqrt{\varepsilon_{r\,1}} \, k_0 \mathbf{r}) &\mathbf{r}\in \Omega_1, \\
\mathbf{M}_{\substack{ e \\ o}   m   n}^{(1)}( \sqrt{\beta_{nl}} \, k_0 \mathbf{r})+ f_{nl} \mathbf{M}_{\substack{ e \\ o}   m   n}^{(2)}(\sqrt{\beta_{nl}}\, k_0 \mathbf{r}) &\mathbf{r}\in \Omega_2,\\
-b_{nl} \mathbf{M}_{\substack{ e \\ o}   m   n}^{(3)}(k_0 \mathbf{r})  &\mathbf{r}\in \Omega_3,
\end{cases}
\label{eq:MagneticModes}
\end{equation}
Dual reasoning leads us to call the eigenfunctions $\CeigB$ associated with the eigenvalues $\beta_{nl}$ {\it magnetic type} modes.
The explicit expression of the coefficients $(a_{nl},\,d_{nl},\,g_{nl})$ and $(b_{nl},\,c_{nl},\,f_{nl})$ in \ref{eq:ElectricModes} and \ref{eq:MagneticModes} are shown in the Appendix \ref{sec:ModesCoef}.

 The functions $\left( \Nii{1}{ \substack{e \\ o} mn}, \Mii{1}{ \substack{e \\ o} mn} \right)$, $\left( \Nii{2}{ \substack{e \\ o} mn}, \Mii{2}{ \substack{e \\ o} mn} \right)$, and $\left( \Nii{3}{ \substack{e \\ o} mn}, \Mii{3}{ \substack{e \\ o} mn} \right)$ are the vector spherical wave functions (VSWFs), whose radial dependence is given by the spherical Bessel functions of the first and second kind, and by the Hankel function of the first kind, respectively \cite{Bohren1998}. The subscripts $e$ and $o$ denote even and odd azimuthal dependence.  The radial mode number $l$ gives the number of maxima along $\hat{\bf r}$ inside the sphere. 

\subsection{Scattered electric field}
The scattered electric field is given by:
\begin{multline}
   {\bf E}_S^{\left(2\right)} \left( {\bf r} \right) = {\bf E}_{S\,0} + \left( \varepsilon_{r\,2} - 1 \right) \times \displaystyle\sum_{m n l} \left( 
 \frac{A_{emnl}}{ \alpha_{nl} - \varepsilon_{r\,2} } \CeigAe +  \right. \\ \left. \frac{A_{omnl}}{ \alpha_{nl} - \varepsilon_{r\,2} } \CeigAo  +   \frac{B_{emnl}}{ \beta_{nl} - \varepsilon_{r\,2} } \CeigBe + \frac{B_{omnl}}{ \beta_{nl} - \varepsilon_{r\,2} } \CeigBo \right),
\label{eq:ExpansionEi}
\end{multline}
where $\displaystyle\sum_{nml} =
 \displaystyle\sum_{n=1}^\infty \, \displaystyle\sum_{m=0}^n \,  \displaystyle\sum_{l=1}^\infty \,$,  
\begin{equation}
\begin{aligned}
A_{ \substack{e \\ o} m n l} &= \frac{\langle 
\CeigA, {\bf \tilde{E}}_i \left( {\bf r} \right) \rangle_{\Omega_2}}{\langle 
\CeigA,\CeigA \rangle_{\Omega_2}}, \\
B_{ \substack{e \\ o} m n l} &= \frac{\langle 
\CeigB, {\bf \tilde E}_i \left( {\bf r} \right) \rangle_{\Omega_2}}{\langle 
\CeigB, \CeigB \rangle_{\Omega_2}},
\end{aligned}
\label{eq:EiCoeff}
\end{equation} 
$\tilde{\mathbf{E}}_i \left( {\bf r} \right)$ is given by Eq. \ref{eq:EincDec}.
  In passive materials where $\im{\varepsilon_{r\,2}} \ge 0$, the quantities $\left| \alpha_{nl} - \varepsilon_{r\,2} \right|$ and $\left| \beta_{nl} - \varepsilon_{r\,2} \right|$ do not vanish as $\omega$ varies because $\im{\alpha_{nl}}<0$ and $\im{\beta_{nl}}<0$. Nevertheless, for any given $\varepsilon_{r\,1}$ and $\eta$, the mode amplitudes $ {A_{ \substack{e \\ o} m n l}}/{ \left( \alpha_{nl} - \varepsilon_{r 2}  \right) }$ and $ {B_{ \substack{e \\ o} m n l}}/{ \left( \beta_{nl} - \varepsilon_{r 2} \right)}$ reach their maximum whenever:
\begin{equation}
\begin{aligned}
     \left|  \alpha_{nl}\left( x, \eta, \varepsilon_{r\,1} \right) - \varepsilon_{r\,2} \left( \omega \right)  \right| &= \underset{x,\omega}{\mbox{min}}; \\ 
     \left| \beta_{nl} \left( x, \eta, \varepsilon_{r\,1} \right) - \varepsilon_{r\,2} \left( \omega \right)  \right| &= \underset{x,\omega}{\mbox{min}},
   \label{eq:Match}
\end{aligned}
\end{equation}
respectively.  These are the resonant conditions for the modes $\CeigA$ and $\CeigB$.

\section{Resonances analysis}

\label{sec:Analysis}
\begin{figure}[!t]
\centering
\includegraphics[width=70mm]{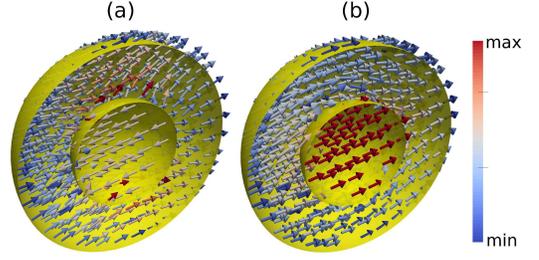}
\caption{Arrow plot of the real part of the {\it bonding} (a) and {\it antibonding} (b) electric dipole eigenmodes of a coated sphere with $\eta=0.5$, $\varepsilon_{r\,1}=4$ and $y=0.1$. Each arrow has the same direction of the eigenmode in the corresponding point of space, while the color of the cone represents its amplitude.}
  \label{fig:ElectricDipoleModes}
\end{figure}

\label{sec:Analysis}

\begin{figure}[!t]
\centering
\includegraphics[width=\linewidth]{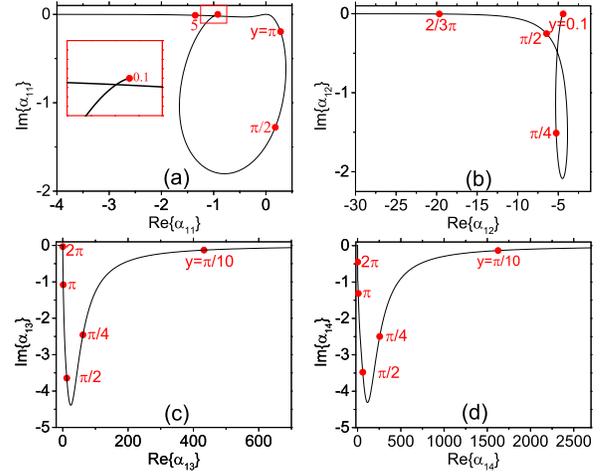}
\caption{Loci spanned in the  complex plane by the eigenvalues $\alpha_{1l}$  associated to the (a) bonding $\left( l=1 \right)$, (b) anti-bonding $\left( l=2 \right)$ electric dipole modes, and to higher order ($l=3,4$) (c-d) electric dipole modes of a coated sphere with $\eta=0.5$ and $\varepsilon_{r\,1}=4$, by varying $y$.}
  \label{fig:LociAlpha1}
\end{figure}

The eigenvalues $\alpha_{nl}$ and $\beta_{nl}$ are independent of the coating's permittivity $\varepsilon_{r\,2}$, they depend on the permittivity of the core $\varepsilon_{r\,1}$,  the aspect ratio $\eta$, and $y$. We now plot the loci they span in the complex plane as a function of $y$ by fixing both $\varepsilon_{r\,1}=4$ and $\eta=0.5$. The loci belong to the half-plane with $\mbox{Im} \left\{ \gamma_s \right\}<0$, as demonstrated in  Eq. \ref{eq:BalanceIm}. The real part of $\gamma_s$ can assume in general both positive and negative values. If $\re{\gamma_s}<0$ the resonant condition \ref{eq:Match} may be satisfied by noble metal coatings in the visible spectral range with $ \mbox{Re} \left\{ \varepsilon_{r\,2} \right\} <0$, giving rise to plasmon resonances (e.g. Ref. \cite{Mayergoyz05}). If $\re{\gamma_s} \ge 0$ the resonant condition $\ref{eq:Match}$ is verified by dielectric coating with $ \mbox{Re} \left\{ \varepsilon_{r\,2} \right\} \ge 0$, giving rise to photonic resonances.
The roots of the two polynomials  are obtained by truncating the power series in Eqs. \ref{eq:PolyP} and \ref{eq:PolyQ} to $h_{max}=50$.

First, we investigate the locus spanned by $\alpha_{11}$, which is shown
 in Fig. \ref{fig:LociAlpha1} (a).  The spatial distribution of the corresponding eigenmodes $\mathbf{C}^{\left( \alpha \right)}_{\substack{ e \\ o}   m 1 1} \left(\mathbf{r} \right)$, shown in Fig. \ref{fig:ElectricDipoleModes}, suggests that these modes can be identified as a {\it bonding} dipole mode. We note that for $y \ll 1$ the eigenvalue $\alpha_{11}$ approaches the value $\alpha_{11}^{\left( 0 \right)}=-0.91523$, in accordance with the electrostatic limit presented in the Appendix \ref{sec:Electrostatic}. This is consistent with Eq. \ref{eq:BalanceRe} that shows that $\re{\gamma_s}<0$ in the quasi-electrostatic limit where $\boldsymbol{\nabla} \times {\bf C}_s \approx {\bf 0} $. By increasing  $y$, both the real and the imaginary part of $\alpha_{11}$  move toward more negative values. For Drude metals with low losses, this fact implies the red shift of the corresponding resonance frequency \cite{maier07}. When $y\approx 1.0$ the quantity $\re{\alpha_{11}}$  reaches a local minimum of $-1.65$ and then starts increasing.
For larger $y$, ${\alpha_{11}}$ moves to the fourth quadrant of the  complex plane, then it further increases until $y\approx 2.13$ where it reaches the global maximum value of $0.383$. Then ${\alpha_{11}}$ passes near the origin of the complex plane in correspondence of $y=4.4836$. This means that it is possible to resonantly excite nanoshell with epsilon-near-zero (ENZ) coatings. This property can be of great use in the flourishing field of ENZ metamaterials \cite{ziolkowski2004propagation,silveirinha2006tunneling} 
especially for enhanced nonlinear generation \cite{Capretti15}. For very large values of $y$,  $\re{\alpha_{11}}$ moves toward minus infinity, asymptotically approaching the negative real axis. It is interesting to note that, as shown in the inset of Fig. \ref{fig:LociAlpha1} (a), there exist two distinct values of $y$, namely $0.32$ and $4.9$ which correspond to the same  eigenvalue ${\alpha}_{11}=-0.99-0.0172 i$. In other words, there exist two coated spheres with the same value of $\varepsilon_{r\,1}$ and $\eta$ but distinct values of $y$ which have the same eigenvalue of the bonding dipolar mode. 

\begin{figure}[!t]
\centering
\includegraphics[width=\linewidth]{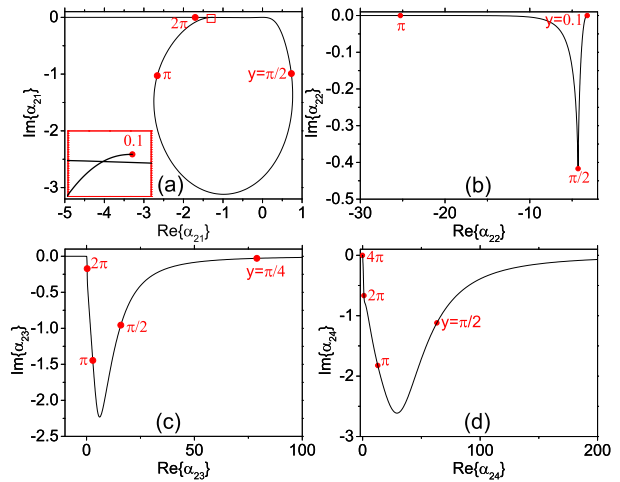}
\caption{Loci spanned in the  complex plane by the eigenvalues $\alpha_{2l}$  associated to the (a) bonding $\left( l=1 \right)$, (b) anti-bonding $\left( l=2 \right)$ electric quadrupole modes, and to higher order ($l=3,4$) (c-d) electric quadrupole modes of a coated sphere with $\eta=0.5$ and $\varepsilon_{r\,1}=4$, by varying $y$.}
  \label{fig:LociAlpha2}
\end{figure}

Next, we consider the eigenvalue $\alpha_{12}$, which is associated to an {\it antibonding} dipole mode, as it is apparent from Fig. \ref{fig:ElectricDipoleModes} (b). We plot in Fig. \ref{fig:LociAlpha1} (b) the locus it spans as $y$ varies. We point out that for $y \ll 1$ the eigenvalue $\alpha_{11}$ approaches the value predicted by the electrostatic theory $\alpha_{12}^{\left( 0 \right)}=-4.37048$. By increasing  $y$, the locus follows a loop, always contained in the third quadrant of the complex plane. Therefore, the antibonding dipole eigenmode can be only resonantly excited if the coating is a metal, namely  $\re{\varepsilon_{r\,2}}<0$, regardless of $y$. For very large values of $y$,  $\re{\alpha_{12}}$ moves toward minus infinity. Also in this case, due to the loop displayed by the locus there exist two distinct values of $y$, namely $0.56$ and $1.40$ which correspond to the same  eigenvalue ${\alpha}_{12}=-5.12 -0.48 i$.

The loci spanned by higher order electric dipole modes $\alpha_{1l}$ with $l=3,4$, shown in Fig. \ref{fig:Figure_1} (c), (d) are instead profoundly different from the ones associated to the bonding and antibonding dipole modes. First, for $y \rightarrow 0$ the real part of $\alpha_{13}, \alpha_{14} \rightarrow \infty$, while $\mbox{Im} \left\{ \alpha_{1l} \right\}$ approaches zero. This fact means that for $y \ll 1$ these modes cannot be practically excited. This is consistent with the theory of electrostatic resonances in nanoshells where these modes do not even exist \cite{mayergoyz2007numerical}.  By increasing $y$, the  values of $\mbox{Re} \left\{ \alpha_{13} \right\}$ and $\mbox{Re} \left\{ \alpha_{14} \right\}$ both move toward smaller values, while the imaginary parts decrease and reach a minimum. Then, $\alpha_{13}$ and $\alpha_{14}$ pass near the origin of the complex plane in correspondence of $y=7.7$ and $y=10.9$, respectively, and eventually move toward minus infinity.

In Figs. \ref{fig:LociAlpha2} we plot the loci spanned by $\alpha_{21}$ and $\alpha_{22}$ of the bonding ($l=1$) and antibonding ($l=2$) electric quadrupole. In this case, for $y \rightarrow 0$ the eigenvalues  $\alpha_{21}$ and $\alpha_{22}$ approach their electrostatic limit  $\alpha_{21}^{\left(0\right)}=-1.24215$ and $\alpha_{22}^{\left(0\right)}=-3.22021$  respectively. Moreover, both loci asymptotically approach the negative real axis for very large values of $y$. Furthermore, $\alpha_{21}$ describes a loop in the complex plane, thus there exist two distinct values of $y$,  which have the same eigenvalue ${\alpha}_{11}=-1.36 -0.0014 i$ associated to a bonding quadrupole mode.  The loci spanned by the eigenvalues $\alpha_{23}$ and $\alpha_{24}$ associated to higher order electric quadrupole modes have the same characteristics of the loci associated to $\alpha_{13}$ and $\alpha_{14}$, which have been already  discussed. 

\label{sec:Analysis}

\begin{figure}[!t]
\centering
\includegraphics[width=\linewidth]{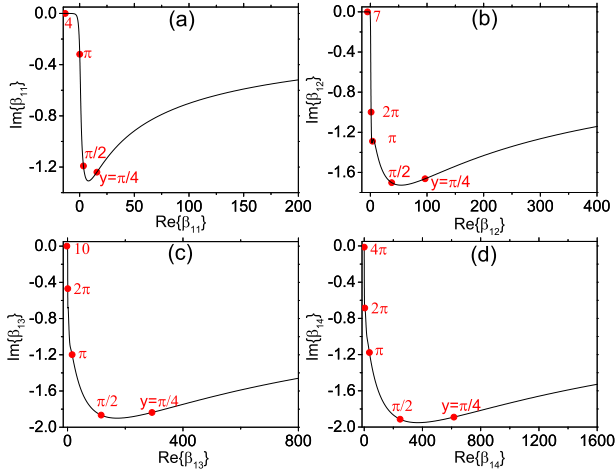}
\caption{Loci spanned in the complex plane by the eigenvalues $\beta_{1l}$ with (a) $l=1$, (b) $l=2$, (c) $l=3$, (d) $l=4$  of the magnetic-type dipole eigenmodes of a coated sphere with $\eta=0.5$ and $\varepsilon_{r\,1}=4$, by varying $y$.}
  \label{fig:LociBeta1}
\end{figure}
\begin{figure}[!t]
\centering
\includegraphics[width=\linewidth]{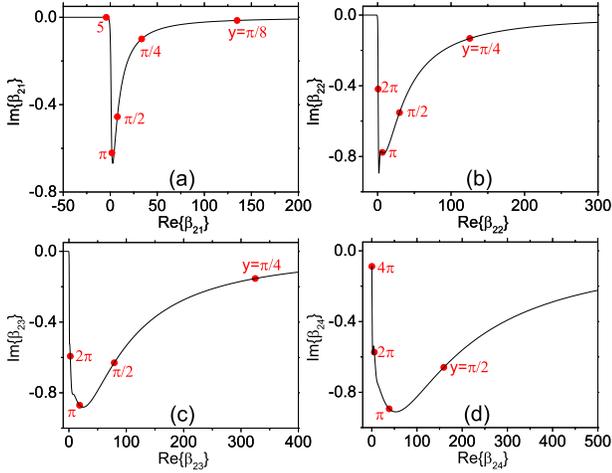}
\caption{Loci spanned in the complex plane by the eigenvalues $\beta_{2l}$ with (a) $l=1$, (b) $l=2$, (c) $l=3$, (d) $l=4$  of the magnetic-type quadrupole eigenmodes of a coated sphere with $\eta=0.5$ and $\varepsilon_{r\,1}=4$, by varying $y$.}
  \label{fig:LociBeta2}
\end{figure}
\begin{figure}[!t]
\centering
\includegraphics[width=70mm]{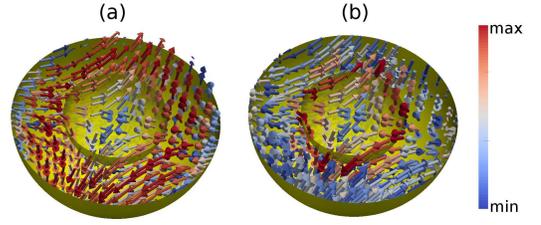}
\caption{Arrow plots of the real part of the magnetic dipole eigenmodes with $n=1$ and $l=1$ (a) and $l=2$ (b) of a coated sphere with $\eta=0.5$, $\varepsilon_{r\,1}=4$ and $y=0.1$. Each arrow has the same direction of the eigenmode in the corresponding point of space, while the color of the cone represents its amplitude.}
  \label{fig:MagneticDipoleModes}
\end{figure}

Let us now consider the loci of the eigenvalues  $\beta_{nl}$, $\forall l=1\ldots 4$ associated to the magnetic dipole ($n=1$) and quadrupole ($n=2$) modes, which are shown in Figs. \ref{fig:LociBeta1} and \ref{fig:LociBeta2}, respectively. 
They all exhibit the same qualitative behaviour of the  eigenvalue of higher order $\left( l \ge 3 \right)$ electric modes. In particular, in the limit for $y \rightarrow 0$ the quantity $\re{\beta_{nl}}$ diverges, thus all the  magnetic modes cannot be practically excited in the electrostatic limit, consistently with the theory of Ref. \cite{mayergoyz2007numerical}. Moreover, by increasing $y$, $\re{\beta_{nl}}$ moves toward smaller values, while the imaginary part decreases and reaches a minimum. Subsequently, they all pass close to the origin of the complex plane and then moves toward minus infinity. It is therefore possible to resonantly excite a magnetic mode in a particle with a metal coating with $\re{\varepsilon_{r\,2}}<0$. We also show in Fig. \ref{fig:MagneticDipoleModes} the magnetic dipole eigenmodes with $n=1$ and $l=1$ (a) and $l=2$ (b) of a coated sphere with $\eta=0.5$, $\varepsilon_{r\,1}=4$ and $y=0.1$.

In conclusion, the only modes that can be resonantly excited in a coated sphere much smaller than the incident wavelength are the bonding and anti-bonding electric ones. In addition, both electric and magnetic eigenvalues asymptotically approach the negative real axis for very large values of $y$. Therefore, in a particle with a metal coating with $\varepsilon_{r\,2}<0$ it is possible to resonantly excite also magnetic modes and higher order electric modes. This result is relevant because in a homogeneous metal sphere with negative permittivity neither magnetic nor higher order electric modes can be resonantly excited. Moreover, the locus of the eigenvalues associated to the bonding (resp. anti-bonding) modes may display a loop, allowing the possibility that two coated sphere with the same value of $\varepsilon_{r\,1}$ and $\eta$ but distinct values of $y$ have the same eigenvalue associated to the same bonding (resp. antibonding) mode. Moreover, all the loci with the exception of the antibonding electric ones, come very close to the origin of the axis. This means that it is possible to resonantly excite nanoshell with epsilon-near-zero (ENZ) coatings.

\subsection{Plane Wave Excitation}
\label{sec:ExpansionPW}

\begin{figure}[!t]
\centering
\includegraphics[width=76mm]{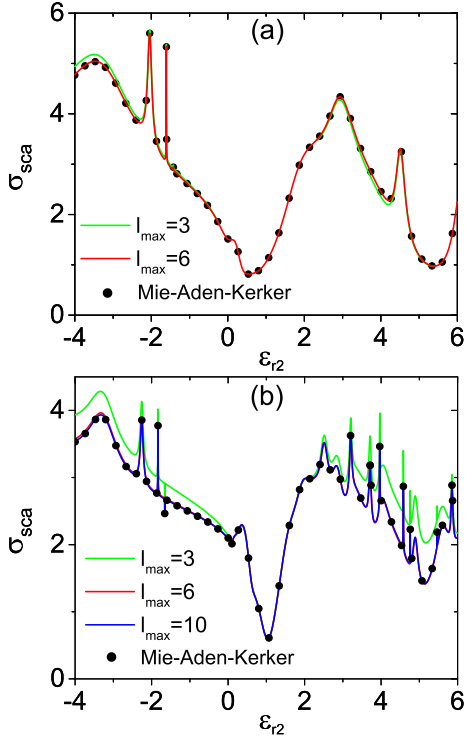}
\caption{Scattering efficiency $\sigma_{sca}$ of a coated sphere with $\varepsilon_{r1}=4$, $y = \pi$ (a), and $y = 2\pi$ (b) excited by a linearly polarized plane wave, as a function of $\varepsilon_{r2} \in \left[-4,6 \right]$ calculated using Eq. assuming $l_{max}= 3$ and $l_{max} = 6$, and $l_{max} = 10$ and with the standard Mie-Aden-Kerker theory. In all the calculations we have assumed $n_{max}=10$.}
  \label{fig:ScatteringEfficiency}
\end{figure}

Let us assume that a $x$-polarized plane wave of unit intensity, propagating along the $z$-axis is exciting the coated sphere. In terms of VSWF the plane wave has the following expression  \cite{Bohren1998} 
\begin{equation}
   {\bf E}_i \left( {\bf r} \right)  =  \displaystyle\sum_{n=1}^\infty  E_n \left[  \Mi{1}{k_0 {\bf r}}{o1n}  - i \Ni{1}{k_0 {\bf r}}{e1n} \right],
  \label{eq:PW_VSWFOmogSphere}
\end{equation}
where
\begin{equation}
E_n =  i^n \frac{2n + 1}{n \left( n+ 1 \right) }.
\label{eq:PWOmogSphere}
\end{equation}
The analytical expression of the electric field $\mathbf{E}_{s\,0}$ scattered by the core in the absence of the coating is provided by the Mie theory \cite{Bohren1998}, $\forall {\bf r} \in \Omega_2 \cup \Omega_3$
\begin{equation}
\label{eq:CampoEs12}
\mathbf{E}_{s\,0}(\mathbf{r}	)=\sum_{n=1}^{\infty} {E}_n \left[ i\,p_n \mathbf{N}_{e1n}^{(3)}(k_0 \mathbf{r})-q_n \mathbf{M}_{o1n}^{(3)}(k_0 \mathbf{r})\right],
\end{equation}
where:
\begin{equation}
p_n=\frac{m_1\psi_n(m_1x)\psi'_n(x)-\psi_n(x)\psi'_n(m_1x)}{m_1\psi_n(m_1x)\xi'_n(x)-\xi_n(x)\psi'_n(m_1x)},\nonumber
\end{equation}
\begin{equation}
q_n=\frac{\psi_n(m_1x)\psi'_n(x)-m_1\psi_n(x)\psi'_n(m_1x)}{\psi_n(m_1x)\xi'_n(x)-m_1\xi_n(x)\psi'_n(m_1x)},
\label{eq:CoefficientiCampoEs12}
\end{equation}
$m_1=\sqrt{\varepsilon_{1\,r}}$, $\psi_n \left( \rho \right) = \rho j_n \left( \rho \right)$ and $\xi_n \left( \rho \right) = \rho h_n^{\left( 1 \right)} \left( \rho \right)$  are the Riccati-Bessel functions. The field scattered by a coated shell excited by a plane wave ${\bf E}_i$ is
\begin{equation}
   \mathbf{E}_{S} = \mathbf{E}_{S\,0} + \tilde{\mathbf{E}}_{S},
   \label{eq:EsTOT}
\end{equation}
where
\begin{multline}
\label{eq:CampoEbar2}
   \tilde{\mathbf{E}}_{S}(\mathbf r)
=  \left( 1 - \varepsilon_{r_2} \right) \times \\ \sum_{n=1}^{\infty}E_n \displaystyle \sum_{l=1}^\infty \left[\frac{ {B}_{nl}}{\varepsilon_{r_2} - \beta_{nl}}\mathbf C_{nl}^{(\beta)}(\mathbf r)
-i\frac{ {A}_{nl}}{\varepsilon_{r_2} - \alpha_{nl}}\mathbf C_{nl}^{(\alpha)}(\mathbf r)\right],
\end{multline}
\begin{align}
A_{nl} &= \frac{\langle \mathbf{C}_{nl}^{(\alpha)}, \mathbf{N}_{e1n}^{(1)} \rangle_{\Omega_2}-p_n\langle \mathbf{C}_{nl}^{(\alpha)}, \mathbf{N}_{e1n}^{(3)} \rangle_{\Omega_2}}{\langle \mathbf{C}_{nl}^{(\alpha)}, \mathbf{C}_{nl}^{(\alpha)} \rangle_{\Omega_2}}, \label{eq:Acoef} \\
B_{nl} &= \frac{\langle \mathbf{C}_{nl}^{(\beta)}, \mathbf{M}_{o1n}^{(1)} \rangle_{\Omega_2}-q_n\langle \mathbf{C}_{nl}^{(\beta)}, \mathbf{M}_{o1n}^{(3)} \rangle_{\Omega_2}}{\langle \mathbf{C}_{nl}^{(\beta)}, \mathbf{C}_{nl}^{(\beta)} \rangle_{\Omega_2}} .
\label{eq:Bcoef}
 \end{align}
In particular, the scattered electric field in the region $\Omega_3$ is given by :
\begin{equation}
\label{eq:CampoE3finale}
\mathbf{E}_S^{\left( 3 \right)}(\mathbf r)=\sum_{n=1}^{\infty} E_n \left[ i\,a_n \mathbf{N}_{e1n}^{(3)}(k_0 \mathbf{r})-b_n \mathbf{M}_{o1n}^{(3)}(k_0 \mathbf{r})\right], \qquad \mathbf{r} \in \Omega_3.
\end{equation}
where
\begin{align}
\label{eq:ancoef}
a_n=&p_n + \left( \varepsilon_{r\,2} - 1 \right) \sum_{l=1}^\infty\frac{{A}_{nl}}{\alpha_{nl} - \varepsilon_{r\,2}} a_{nl},\\
\label{eq:bncoef}
b_n=&q_n +  \left( \varepsilon_{r\,2} - 1 \right) \sum_{l=1}^\infty \frac{ {B}_{nl}}{\beta_{nl} - \varepsilon_{r\,2} } b_{nl}.
\end{align}
The coefficients $a_{nl}$ and $b_{nl}$ have been introduced in the Eqs. \ref{eq:ElectricModes} and their expression is shown in the Appendix \ref{sec:ModesCoef}.
In the framework of the proposed modal expansion, we now  calculate the scattering efficiency $\sigma_{sca}$ of a coated sphere \cite{Bohren1998}, when it is excited by a linearly polarized plane wave:
\begin{equation}
\sigma_{sca} = \frac{C_{sca}}{\pi y^2} =\frac{2}{y^2}\sum_{n=1}^{\infty}(2n+1)\left(|a_n|^2+|b_n|^2\right),
\label{eq:SigmaSca}
\end{equation}
where $a_{n}$ and $b_{n}$ are given by Eqs. \ref{eq:ancoef}-\ref{eq:bncoef}.
In Fig. \ref{fig:ScatteringEfficiency} (a)  we  plot $\sigma_{sca}$ for the same sphere considered in the previous section ($\eta=0.5$, $\varepsilon_{r\,1}=4$), and with two different values of $y$, namely $\pi$ and $2\pi$ as a function of a real permittivity $\varepsilon_{r\,2} \in \left[-4,6 \right]$, calculated by truncating the exterior sum of Eq. \ref{eq:SigmaSca} to $n_{max}=10$, and the inner sum to $l_{max}=3$ (blue line) and to $l_{max}=6$ and $l_{max}=10$. We compare them with the standard Mie-Aden-Kerker solution \cite{Bohren1998} calculated assuming the same value of $n_{max}$. In the case of  $y=\pi$, shown in panel (a), the agreement is already good when $l_{max}=3$. When $y$ is increased to $2\pi$, for $l_{max}=3$ there is a moderate disagreement with the Mie-Aden-Kerker theory, for $l_{max}=6$ the outcomes of the two approaches become almost indistinguishable. 


\section{Design of electromagnetic cloaks}
\label{sec:Design}

In this section, we use the introduced approach to design the permittivity of the coating of an homogeneous sphere of assigned size and material composition to achieve several goals, namely the cancellation of the backscattering, the zeroing of a prescribed scattering order, and the maximization of the magnitude of a field component in a given point of space. We will show that, within the proposed framework, the fulfilment of these goals requires one to only find the roots of a polynomial equation.

\subsection{Backscattering Cancellation}

\begin{figure}[!t]
\centering
\includegraphics[width=70mm]{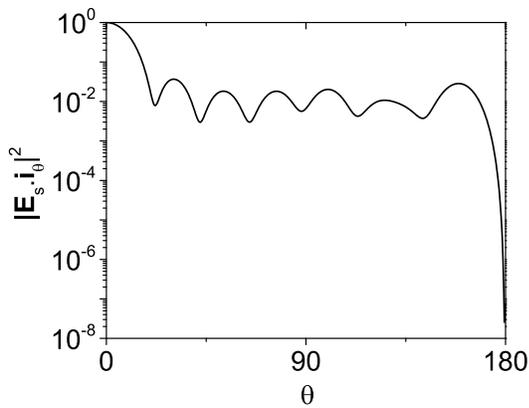}
\caption{Radiation diagram for $\phi = 0$ as a function of the angle $\theta$ for the a coated sphere with $y=2\pi$, $\varepsilon_{r\,1}=4$, and with a value of $\varepsilon_{r\,2}$ designed to enforce a vanishing back-scattering.}
  \label{fig:BackScattering}
\end{figure}

More than three decades ago Kerker et al. first demonstrated the suppression of the back-scattering in magneto-dielectric spheres of arbitrary size with $\varepsilon = \mu$ \cite{kerker1983electromagnetic}. More recently, Nieto et al. \cite{nieto2011angle} predicted that, when the scattering response of a {\it small} non-magnetic sphere is dominated by the magnetic and electric dipoles multipolar order, vanishing backscattering can result from their destructive interference. This scenario, that generalizes the Kerker's condition, has been experimentally observed both in the microwaves \cite{geffrin2012magnetic} and in the visible spectral range \cite{fu2013directional,Person13}. An additional extension of the Kerker's conditions that describes the suppression of the backscattering from a sphere when excited by a local dipole source has also been introduced in Ref. \cite{Rolly:12}. Furthermore, the generalized Kerker condition has been also verified in subwavelength metal-dielectric core-shell particles \cite{Liu12}, core shell nanowires \cite{liu2013scattering}, and silicon nanodisks \cite{Staude13}, and to particles with cylindrical symmetry \cite{Zambrana:13}. It is also worth to point out that the backscattering cancellation from a dielectric sphere is also possible  even when the size of the particle is comparable with the incident wavelength and many scattering orders are involved, as shown in Ref. \cite{Forestiere16}. In this section, we introduce a procedure to cancel the backscattering of a homogeneous sphere by designing the permittivity of its coating. We assume that the sphere has inner radius $R_1=\lambda/2$, i.e. $x=\pi$, outer radius $R_2=\lambda$, i.e. $y=2\pi$,  and is excited by a $x$-polarized plane wave of unit intensity, propagating along the $z$-axis. Within the framework of the proposed approach, the determination of the  permittivities of the coating that cancel the backscattering of the coated sphere only requires one to find the roots of a polynomial equation. The radiation pattern is defined as
\begin{equation}
 \mathbf{E}_S^\infty \left( \theta, \phi, \varepsilon_{r2} \right) = \displaystyle\lim_{r \rightarrow \infty}  \left[ r{e^{- i k_0 r}} {\bf E}_S^{\left(3\right)} \right],
\end{equation} 
where $\theta$ and $\phi$ are the polar and azimuthal angles, respectively.
Due to symmetry considerations the only non-vanishing component of the radiation pattern in the backscattering direction ($\theta = \pi$) is $\mathbf{E}_S^\infty \cdot {\bf i}_\theta$. Therefore, we have to find the zeros of $\mathbf{E}_S^\infty \cdot {\bf i}_\theta$ as a function of $\varepsilon_{r2}$, where $\mathbf{E}_S^\infty \cdot {\bf i}_\theta$ is expressed as:

\begin{equation}
\label{eq:diffSca}
   {\bf E}_S^\infty \cdot {\bf i}_\theta =\sum_{n=1}^{\infty} E_n \left[ i\,a_n \Ni{\infty}{\theta,\phi}{e1n}  \cdot {\bf i}_\theta -b_n 
\Mi{\infty}{\theta,\phi}{o1n} \cdot {\bf i}_\theta   
   \right],
\end{equation}
where $a_{n}$ and $b_{n}$ are defined in Eq. \ref{eq:ancoef}-\ref{eq:bncoef}, $
 \Mii{\infty}{o1n} = \displaystyle\lim_{  r \rightarrow \infty}  \left[ k_0 {r} {e^{- i k_0 r}} \Mii{3}{o1n} \right]
$,
$
 \Nii{\infty}{e1n} = \displaystyle\lim_{r \rightarrow \infty}  \left[ k_0 {r} {e^{- i k_0 r}} \Nii{3}{e1n} \right]
$.
We set $y=2\pi$, $\eta=0.5$, $\theta=\pi$ and $\phi=0$ in the expression \ref{eq:diffSca} truncated with $n_{max}=10$ and $l_{max}=10$. Then, we substitute Eqs.  \ref{eq:ancoef}-\ref{eq:bncoef} into Eq. \ref{eq:diffSca}, and we put all the terms in the resulting sum over a common denominator, obtaining in this way a rational function and we zero the resulting numerator, which is a polynomial in $\varepsilon_{r\,2}$. Among the different solutions, we chose $\varepsilon_{r\,2} = -2.2756023 + 0.0840900 i$. To validate this result, we plot in Fig. \ref{fig:BackScattering} (b) the squared magnitude of the radiation pattern of the designed sphere as a function of the angle $\theta$ for $\phi=0$, computed by using the Mie-Aden-Kerker solution with $n_{max}=10$. We achieved a ratio between the back- and the forward- scattered power of -36dB. 

It is worth noting that the achieved backscattering suppression cannot be attributed to the interference of solely electric and magnetic dipoles as in \cite{kerker1983electromagnetic,nieto2011angle,fu2013directional,Person13}, but originates from a complex interplay of many electric and magnetic scattering orders, which are significant up to $n=10$.

 In conclusion, our method enables the fine engineering of the zeros of the radiation diagram of a nanosphere, through the design of the permittivity of its coating. In particular, in our example we designed a coated particle with a pronounced anisotropy of its scattering response, where the forward scattering strongly dominates over the backscattering. We envisage that the algorithm outlined in this section will facilitate the engineering of highly directional metal or dielectric nanoantennas.  

\subsection{Scattering order suppression}

In 1975, Kerker demonstrated that nonabsorbing coated concentric spheres \cite{Chew:76} or ellipsoids \cite{Kerker:75} composed of an inner ellipsoidal region and an outer confocal ellipsoidal shell, feature zero scattering for certain combinations of dielectric constants, thus behaving as invisible objects. Later, the design of invisibility was further investigated in Ref. \cite{Alu05} by using with plasmonic and metamaterial coatings. An algorithm for cancelling the scattering from an arbitrarily shaped coated object in the limit of small particle has been recently proposed in \cite{forestiere2014cloaking}. It is worth noting that all the aforementioned approaches cancel solely the dipole scattering order and hold true only in the small particle regime.

In the following, we further generalize these results, showing how to suppress a prescribed electric or magnetic multipolar order scattered by a given sphere of any size by cloaking it with a homogeneous coating. This is accomplished by zeroing the corresponding scattering coefficient $a_n$ or $b_n$, given in Eq. \ref{eq:ancoef}, \ref{eq:bncoef}, which in our representation can be recast as a rational function of $\varepsilon_{r\,2}$. 
\begin{figure}[!t]
\centering
\includegraphics[width=\linewidth]{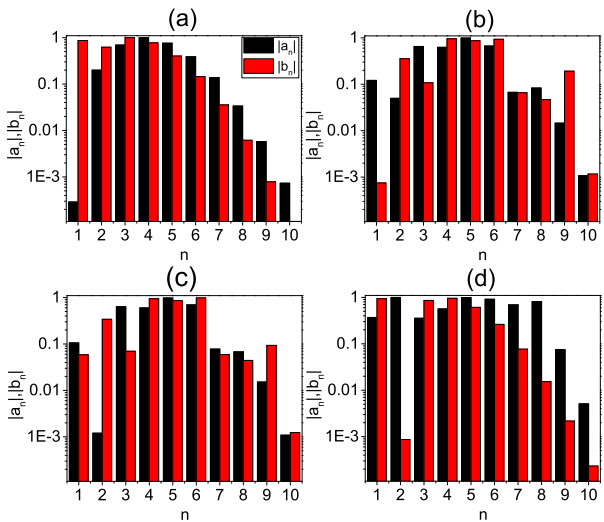}
\caption{Magnitude of the scattering coefficients $a_n$ and $b_n$ of a coated sphere with core permittivity $\varepsilon_{r1}=4$, aspect ratio $\eta=0.5$, and $y = 2 \pi$ whose coating permittivity $\varepsilon_{r2}$ was designed to cancel the electric dipole (a), the magnetic dipole (b), the electric quadrupole (c), or the magnetic quadrupole (d). The used values of $\varepsilon_{r2}$ are listed in Tab. \ref{tab:MultipoleCancellation}.}
  \label{fig:ScattOrdSuppression}
\end{figure}
First, we recast all the terms in the sum of Eqs. \ref{eq:ancoef}, \ref{eq:bncoef} over a common denominator, obtaining in this way a rational function and we zero the resulting numerator, which is a polynomial in $\varepsilon_{r\,2}$. 
The values of permittivity $\varepsilon_{r2}$ 
 that suppress the electric dipole, magnetic dipole,  electric quadrupole, magnetic quadrupole scattering order are listed in table \ref{tab:MultipoleCancellation}. To validate these results, in Fig. \ref{fig:ScattOrdSuppression} we plot the magnitude of the scattering coefficients $a_n$ and $b_n$ of the four designed coated spheres. The scattering orders have been calculated by using the Mie-Aden-Kerker solution \cite{Bohren1998,aden1951scattering}. We note that in each scenario the suppressed scattering order is roughly three orders of magnitudes smaller than the dominant one. We also point out that there is a residual multipolar scattering because we only considered a finite number of radial eigenmodes ($l_{max}=10$).

\begin{table}[htbp]
\centering
\caption{\bf Values of permittivity of the coating suppressing a given scattering order}
\begin{tabular}{ccc}
\hline
$\varepsilon_{r2}$ & Electric & Magnetic \\
\hline
Dipole     & $0.10307+3.171 \cdot 10^{-6} i$ & $+3.99954+1.432 \cdot 10^{-4} i$ \\
Quadrupole & $4.03763+7.769 \cdot 10^{-4} i$ & $-2.99455+0.0110425 i$ \\
\hline
\end{tabular}
  \label{tab:MultipoleCancellation}
\end{table}

\subsection{Field Maximization}

Nanoantennas are optical devices which efficiently couple the incoming electromagnetic radiation to modes localized in regions with dimensions well below the diffraction limit \cite{novotny2006}. In the last decade, metal nanoantennas have been proposed for many technological applications \cite{schuller2010}. More recently, it became apparent that metal nanostructures are plagued by high losses \cite{khurgin2015deal} which prevent them from becoming commercial devices, and therefore dielectric nanoantennas have been proposed as a suitable low-loss alternative \cite{Krasnok:12}. 

Thus, it is crucial to rationally design metal or dielectric nanostructures capable of producing the highest field enhancement at well defined locations and targeted frequency spectra for device applications. Heuristic approaches to the design of metallic nanostructures featuring high field enhancement relied on self-similar chains of metal nanospheres \cite{Li2003,Dai2008}. In addition, enhanced fields can be achieved by introducing a small gap in the metal structure \cite{Schuck05} or exploiting the lightning-rod effect taking place at a sharp metal tip \cite{Gersten80}. More recently, optimization algorithms have been also employed to maximize the field enhancement  \cite{forestiere10,forestiere12,feichtner12,forestiere15}.

\begin{figure}[!t]
\centering
\includegraphics[width=\linewidth]{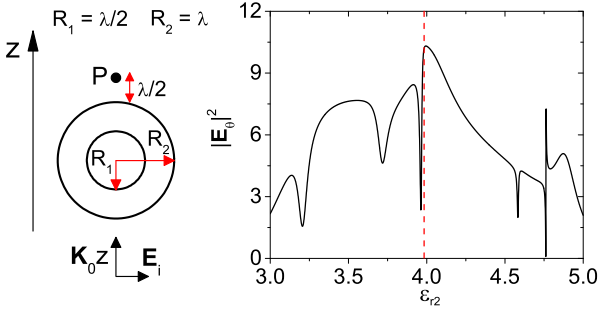}
\caption{(a) Squared magnitude in the point $\left( 0, 0, h \right)$ of the component $E_{\theta}$ of the field scattered by a coated sphere with $y=2\pi$, $\eta=0.5$, $\varepsilon_{r\,1}=4$ as a function of the permittivity of the coating $\varepsilon_{r\,2}$. With a vertical dashed line the designed value of $\varepsilon_{r\,2}$ that guarantees the maximum value of $\left|E_{\theta} \right|^2$.}
  \label{fig:Max}
\end{figure}

We now use the spectral framework developed so far to design the permittivity of the coating that locally maximizes a component of the electric field in a given point of space. In particular, we maximize the squared magnitude of the $\hat{\theta}$ component of the electric field scattered by the coated sphere at the point $\left(x,y,z\right) = \left(0,0, z_0 \right)$, as shown in the sketch of Fig. \ref{fig:Max}. Only in this case, we assumed $\varepsilon_{r\,2}$ to be real. Thus, starting from Eqs. \ref{eq:EsTOT},\ref{eq:CampoEbar2} and truncating them to $n_{max}=10$ and $l_{max}=8$, we calculate the derivative $\frac{d\left\| \mathbf{E}_\theta \right\|^2}{d \varepsilon_{r\,2}}$. We put all the resulting terms over a common denominator obtaining in this way a rational function and we zero the resulting numerator, which is a polynomial in $\varepsilon_{r\,2}$. We set the following parameters $y=2\pi$, $\eta=0.5$ and $z_0=3/2\lambda$. Among the different solutions, we choose $\varepsilon_{r\,2} = 3.9833$. In Fig. \ref{fig:Max} we plot the magnitude  of the component $E_{\theta}$ of the field scattered by a coated sphere with $y=2\pi$, $\eta=0.5$, $\varepsilon_{r\,1}=4$ in the point $\left( 0, 0, h \right)$ as a function of the permittivity of the coating $\varepsilon_{r\,2}$. With a vertical dashed line the designed value of $\varepsilon_{r\,2}$ that guarantees the maximum value of $\left|E_{\theta} \right|^2$. This plot validates our maximization.

\section{Conclusions}
We introduced an alternative representation of the electromagnetic field scattered from a homogeneous sphere coated with a homogeneous layer of uniform thickness. Specifically, we represented the electromagnetic field in terms of a set of eigenfunctions of an auxiliary eigenvalue problem, which are independent of the permittivity of the coating. We used this theory for the analysis of the resonances of core-shell particles, by plotting the loci of its electric and magnetic eigenvalues as a function of the size parameter.  Furthermore, to illustrate the great potential of this method, we design the permittivity of the coating to zero the backscattering, to zero a prescribed multipolar order of the scattered field, and to maximize the electric field in a given point of space.

\appendix
\section{Characteristic Polynomial Coefficients}
\label{sec:PolyCoef}
In this section we show the analytical expressions of the coefficients needed to calculate the polynomials in Eqs. \ref{eq:PolyP},\ref{eq:PolyQ}:
\begin{equation}
\sigma_{hk}=\frac{(-1)^{h+n} 2^{-h}  x^{-k} y^{-k}}{k! (h-k)!},
\end{equation}
\begin{multline}
r_{hk}^{\left(n\right)} =y^h x^{2 k} j_{k-n-1}(x) \left( \left( m_1 x\right) j_{n}(m_1 x) \right)' \times \\ \left[y h_n^{(1)}(y) j_{h-k+n+1}(y)+\left( y h_{n}^{(1)}(y)\right)' j_{h-k+n}(y)\right] \\ -x^h y^{2 k} j_{k-n-1}(y) \left(y h_{n}^{(1)}(y)\right)' \times \\     \left[ m_1^2 x j_n(m_1 x) j_{h-k+n+1}(x) +\left(  \left( m_1 x\right) j_{n}(m_1 x) \right)' j_{h-k+n}(x)\right],
\end{multline}

\begin{multline}
s_{hk}^{\left(n\right)} = x^h y^{2 k+1} j_{k-n-2}(y) h_n^{(1)}(y) \times \\
\left[ \left( m_1^2 x \right) j_{n} \left(m_1 x \right) j_{h-k+n+1}(x) + \left( \left( m_1 x \right) j_{n} \left(m_1 x \right)\right)' j_{h-k+n}(x)\right]\\-y^h x^{2 k +1} m_1^2 j_{k-n-2}(x) j_n(m_1 x) \times  \\ \left[y h_n^{(1)}(y) j_{h-k+n+1}(y)\right. 
\left.+\left(  y h_{n}^{(1)}(y)\right)' j_{h-k+n}(y)\right]+ \\ 
\left( n + 1 \right) j_n(m_1 x) m_1^2  h_n^{(1)}(y) \times \\
\left[x^{h+1} y^{2 k} j_{h-k+n+1}(x) j_{k-n-1}(y)  - y^{h+1} x^{2 k} j_{h-k+n+1}(y)  j_{k-n-1}(x) \right] +\\ 
\left( n + 1 \right) \left[ x^h y^{2 k} j_{h-k+n}(x)j_{k-n-1}(y) - y^h x^{2 k} j_{h-k+n}(y)j_{k-n-1}(x) \right] \times \\
 \left[ m_1^2 j_n(m_1 x) \left( y h_n^{(1)}(y) \right)' + h_n^{(1)}(y) \left( \left(m_1 x \right) j_n(m_1 x) \right)'   \right],
\end{multline}
\begin{multline}
t_{hk}^{\left(n\right)}=-m_1^2 (n+1) h_n^{(1)}(y) j_n(m_1 x)  \times \\ 
\left\{x^h y^{2 k} \left[ y j_{k-n-2}(y)+(n+1) j_{k-n-1}(y)\right] j_{h-k+n}(x)\right. \\
\left.-y^h x^{2 k} \left[x j_{k-n-2}(x)+(n+1) j_{k-n-1}(x) \right] j_{h-k+n}(y) \right\},
\end{multline}

\begin{multline}
u_{hk}^{\left(n\right)} =x^h y^{2 k} h_n^{(1)}(y) j_{k+n+1}(y) \times \\ \left[j_n(m_1 x) j_{h-k-n-2}(x)+m_1 j_{n+1}(m_1 x) j_{h-k-n-1}(x)\right]\\
-y^h x^{2 k} j_{k+n+1}(x) j_n(m_1 x) \times \\ \left[h_n^{(1)}(y) j_{h-k-n-2}(y)+h_{n+1}^{(1)}(y) j_{h-k-n-1}(y)\right],
\end{multline}
\begin{multline}
v_{hk}^{\left(n\right)} =m_1 y^h x^{2 k} j_{k+n}(x) j_{n+1}(m_1 x) \times \\ \left[h_n^{(1)}(y) j_{h-k-n-2}(y)+h_{n+1}^{(1)}(y) j_{h-k-n-1}(y)\right]\\
-x^h y^{2 k} h_{n+1}^{(1)}(y) j_{k+n}(y) \times \\ \left[j_n(m_1 x) j_{h-k-n-2}(x)+m_1 j_{n+1}(m_1 x) j_{h-k-n-1}(x)\right],
\end{multline}
where $j_n$ and $h_n$ are the spherical Bessel and Hankel functions of the first kind, respectively.
\section{Eigenmodes Coefficients}
In this section we provide the analytical expressions of the coefficients needed to calculate the electric and magnetic modes of \ref{eq:ElectricModes},\ref{eq:MagneticModes}:
\label{sec:ModesCoef}

\begin{equation}
\small
\begin{aligned}
a_{nl}&= \frac{\psi_n(\sqrt{\alpha_{nl}} y) \left[\sqrt{\alpha_{nl}} \chi _n(\sqrt{\alpha_{nl}} x) \psi'_n(m_1 x)-m_1 \chi'_n(\sqrt{\alpha_{nl}} x) \psi _n(m_1 x) \right]}{\xi _n(y) \left[m_1 \chi'_n(\sqrt{\alpha_{nl}} x) \psi _n(m_1 x)-\sqrt{\alpha_{nl}} \chi _n(\sqrt{\alpha_{nl}} x) \psi'_n(m_1 x)\right]}\\
&+\frac{\chi _n(\sqrt{\alpha_{nl}} y) \left[m_1 \psi _n(m_1 x) \psi'_n(\sqrt{\alpha_{nl}} x)- \sqrt{\alpha_{nl}} \psi _n(\sqrt{\alpha_{nl}} x) \psi'_n(m_1 x)\right]}{\xi _n(y) \left[m_1 \chi'_n(\sqrt{\alpha_{nl}} x) \psi _n(m_1 x)-\sqrt{\alpha_{nl}} \chi _n(\sqrt{\alpha_{nl}} x) \psi'_n(m_1 x)\right]} \\
d_{nl}&=\frac{m_1 \left[\chi'_n(\sqrt{\alpha_{nl}} x) \psi _n(\sqrt{\alpha_{nl}} x)-\chi _n(\sqrt{\alpha_{nl}} x) \psi'_n(\sqrt{\alpha_{nl}} x)\right]}{m_1 \chi'_n(\sqrt{\alpha_{nl}} x) \psi _n(m_1 x)-\sqrt{\alpha_{nl}} \chi _n(\sqrt{\alpha_{nl}} x) \psi'_n(m_1 x)}, \\
 g_{nl} &= \frac{m_1 \psi _n(m_1 x) \psi'_n(\sqrt{\alpha_{nl}} x)-\sqrt{\alpha_{nl}} \psi _n(\sqrt{\alpha_{nl}} x) \psi'_n(m_1 x) }{ m_1 \psi_n(m_1 x) \chi'_n(\sqrt{\alpha_{nl}} x) -\sqrt{\alpha_{nl}} \chi _n(\sqrt{\alpha_{nl}} x) \psi'_n(m_1 x)}
\end{aligned}
\end{equation}

\begin{equation}
\small
\begin{aligned}
b_{nl}&= \frac{
\psi _n(\sqrt{\beta_{nl}} y) \left[
m_1 \chi _n(\sqrt{\beta_{nl}} x) \psi'_n(m_1 x) -\sqrt{\beta_{nl}} \chi'_n(\sqrt{\beta_{nl}} x) \psi _n(m_1 x) \right]}
{\sqrt{\beta_{nl}} \xi _n(y) \left[\sqrt{\beta_{nl}} \chi'_n(\sqrt{\beta_{nl}} x) \psi _n(m_1 x)-m_1 \chi _n(\sqrt{\beta_{nl}} x) \psi'_n(m_1 x)\right]}\\
&+\frac{\chi _n(\sqrt{\beta_{nl}} y) \left[\sqrt{\beta_{nl}} \psi _n(m_1 x) \psi'_n(\sqrt{\beta_{nl}} x)-m_1 \psi _n(\sqrt{\beta_{nl}} x) \psi'_n(m_1 x)\right]}{\sqrt{\beta_{nl}} \xi _n(y) \left[\sqrt{\beta_{nl}} \chi'_n(\sqrt{\beta_{nl}} x) \psi _n(m_1 x)-m_1 \chi _n(\sqrt{\beta_{nl}} x) \psi'_n(m_1 x)\right]} \\
 c_{nl}&=\frac{m_1 \chi'_n(\sqrt{\beta_{nl}} x) \psi _n(\sqrt{\beta_{nl}} x)-m_1 \chi _n(\sqrt{\beta_{nl}} x) \psi'_n(\sqrt{\beta_{nl}} x)}{\sqrt{\beta_{nl}} \chi'_n(\sqrt{\beta_{nl}} x) \psi _n(m_1 x)-m_1 \chi _n(\sqrt{\beta_{nl}} x) \psi'_n(m_1 x)}, \\
f_{nl} &= \frac{m_1 \psi _n(\sqrt{\beta_{nl}} x) \psi'_n(m_1 x)-\sqrt{\beta_{nl}} \psi _n(m_1 x) \psi'_n(\sqrt{\beta_{nl}} x)}{m_1 \chi _n(\sqrt{\beta_{nl}} x) \psi'_n(m_1 x)-\sqrt{\beta_{nl}} \chi'_n(\sqrt{\beta_{nl}} x) \psi _n(m_1 x)} 
\end{aligned}
\end{equation}
where $\psi_n \left( \rho \right) = \rho j_n \left( \rho \right)$, $\chi_n \left( \rho \right) = -\rho y_n \left( \rho \right)$, and $\xi_n \left( \rho \right) = \rho h_n^{\left( 1 \right)} \left( \rho \right)$  are the Riccati-Bessel functions.

\section{Electrostatic Limit}
\label{sec:Electrostatic}
A coated sphere with aspect ratio $\eta = \frac{R_1}{R_2}$ and core's permittivity $\varepsilon_{r\,1}$, features in the the electrostatic limit two resonant eigenvalues $\alpha^{\left(0\right)}_{n\,1}$ and $\alpha^{\left(0\right)}_{n\,2}$, which are solution of the following second order equation:
\begin{equation}
\small
 \left(\alpha^{\left(0\right)}_{n} \right)^2 + \frac{ \left( \varepsilon_{r\,1} +1\right) n \left( n + 1\right) \eta^{3} + \varepsilon_{r\,1} n^2   + \left( n + 1 \right)^2 }{ n \left( n + 1 \right) \left( 1 - \eta^{3} \right) } \left(\alpha^{\left(0\right)}_{n} \right) + \varepsilon_{r\,1} = 0
\end{equation}
Each eigenvalue is $2n + 1$ degenerate because of the spherical symmetry.


\end{document}